\documentclass[12pt]{iopart}
\usepackage{graphicx}
\begin{document}
\title{High coherence photon pair source for quantum communication}
\author{Matth\"{a}us Halder$^1$, Alexios Beveratos$^2$, Rob T Thew$^1$, Corentin Jorel$^1$, Hugo Zbinden$^1$ and Nicolas~Gisin$^1$}
\address{$^1$ Group of Applied Physics, University of Geneva, 1211
Geneva 4, Switzerland}
\address{$^2$ LPN-CNRS Route de Nozay 91460 Marcoussis, France}
\ead{matthaeus.halder@physics.unige.ch}

\begin{abstract}
This paper reports a novel single mode source of narrow-band
entangled photon pairs at telecom wavelengths under continuous
wave excitation, based on parametric down conversion. For only
7\,mW of pump power it has a created spectral radiance of
0.08~pairs per coherence length and a bandwidth of 10\,pm
(1.2\,GHz). The effectively emitted spectral brightness reaches
$3.9*10^5$ pairs $s^{-1}$ pm$^{-1}$. Furthermore, when combined
with low jitter single photon detectors, such sources allow for
the implementation of quantum communication protocols without any
active synchronization or path length stabilization. A HOM-Dip
with photons from two autonomous CW sources has been realized
demonstrating the setup's stability and performance.

\end{abstract}

\submitto{\NJP}

\section{Introduction}

In quantum communication, flying qubits are encoded either on
single photons, or pairs of entangled photons. The latter are a
key element for remote communication protocols like teleportation
or quantum repeaters. Such protocols are based on a local joint
measurement of two photons (e.g. Bell state measurement, BSM)
originating from separated sources and rely on the fact that the
photons to be measured are indistinguishable. To provide temporal
indistinguishability of independent photons, their timing
precision has to be better than their coherence time. For photons
originating from different remote sources, this can be achieved by
pulsed emission times, synchronized by an external clock
\cite{Zeilinger, swap06}. Major drawbacks are the need for
accurate synchronization of the lasers as well as precise matching
and stabilization of the optical path lengths.

Alternatively, timing can be obtained by postselecting appropriate
photon pairs by detection \cite{swap93}. To allow for precise
timing, the coherence time of the photons has to be longer than
the temporal resolution of the detectors. As an advantage, the
sources do not require any synchronization and hence allow the use
of relatively simple continuous wave (CW) excitation. In addition
due to the long coherence length, the setup is less sensitive to
path length fluctuations which is a limiting factor for joint
measurements under pulsed excitation.

Entangled photon pairs may, for example, be produced by bi-exciton
cascade emission of quantum dots \cite{benson, shields} or
intracavity atomic ensembles \cite{thompson}, but due to their
relatively early development stage, these techniques are still not
practical for quantum communication. Alternatively nonlinear
effects like spontaneous 4-wave mixing \cite{kumar07, rarty07,
migdal, takesue05} or spontaneous parametric down conversion
(SPDC) in nonlinear crystals \cite{Zeilinger, karlson, tanzilli,
huber, parisCounterpropagate} can be used. These systems appear to
be more practical, but have the disadvantage of a broad emission
spectrum (from one to tens of nm), providing an insufficient
coherence length to tolerate length fluctuations in optical fibres
($4*10^{-6}~K^{-1}$) of several kilometers.

A narrow-bandwidth emission spectrum can be obtained by counter
propagating SPDC \cite{parisCounterpropagate} or SPDC in photonic
crystals \cite{photonicCrystal}, but are still under development
and efficiencies are extremely low. Another approach to achieving
a long coherence time consists in either inserting the nonlinear
crystal in a cavity \cite{polzik} or using very narrow bandpass
filters \cite{HOM25}. The first technique generally provides a
smaller bandwidth whereas the latter has the advantage of easy
maintenance, advantageous e.g. for field experiments. In the first
part of this paper we present a CW source of entangled photon
pairs with long coherence times. This is achieved by combining the
high conversion efficiency of SPDC in a nonlinear PPLN waveguide
with narrow bandwidth filtering via a phase shifted fibre Bragg
grating (PSFBG). In the second part, we describe the realization
and implementation of different detection techniques.

Fibre optical networks provide an existing resource for long
distance communication. To take advantage of this and in order to
minimize propagation losses, light at telecommunication
wavelengths has to be used. At this wavelength -1560\,nm in our
case - single photons are usually detected by InGaAs avalanche
photo detectors (APDs), which provide timing resolution of up to
100\,ps and quantum efficiencies \cite{InGaAs} of up to 30\%, but
generally, only in gated mode. This makes them unsuitable for
combination with CW-sources due to the lack of synchronization
signals. In order to meet the strict conditions on the detectors
for high timing resolution and free running operation, we focus on
two new generation detectors which have been developed recently.
We implemented mid-infrared single photon detection by
up-conversion combined with Si-APDs \cite{robUp, yamamotoUp} as
well as superconducting detectors provided by SCONTEL \cite{sspd}.

To demonstrate the performance of the sources in combination with
high resolution detectors, we present a Hong-Ou-Mandel (HOM) dip
in the last part of the paper. The results are discussed in the
conclusion.

\section{The photon pair source}

\subsection{Entanglement by parametric down conversion}

In the following, the different components of our source are
described in detail. Figure~\ref{setup} shows a schematic setup of
our narrow band CW source. Pairs of energy-time entangled photons
are created by SPDC in a nonlinear crystal pumped at
$\lambda_p=780\,nm$. The crystal is a $50\,mm$-long periodically
poled Lithium Niobate (PPLN) wave\-guide (HC Photonics). Phase
matching is such that SPDC produces degenerate pairs of signal and
idler photons at 1560\,nm; with a spectral distribution of
$\Delta\lambda_0 = 80\,nm$. The crystal's temperature can be tuned
and the nonlinear conversion efficiency was measured to be
$10^{-5}$. Through energy conservation, signal and idler photons
satisfy the relationship
$\lambda_p^{-1}=\lambda_s^{-1}+\lambda_i^{-1}$, hence detecting
the signal photon at $\lambda_s$ projects the corresponding idler
photon onto $\lambda_i$. The created photons have the same
polarization and exit the PPLN waveguide collinearly. They are
coupled into a standard optical single mode fibre with an
efficiency of ~30\%. A bulk high-pass Silicon filter (Si) is
placed just before coupling into the fibre in order to block the
remaining pump light while transmitting ($T=90\%$) the created
photons.

\subsection{Long coherence time by narrow filtering}

\begin{figure}[h!]
\begin{center}
\includegraphics[width=0.7\columnwidth]{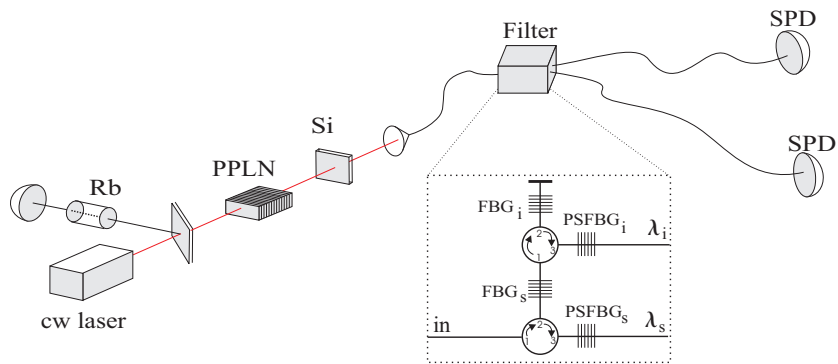}
\caption {\textit{Setup of the source. A continuous laser (CW),
stabilized on the $D_2$-transition line of Rubidium (Rb), is
pumping a nonlinear PPLN crystal. Pairs of entangled signal and
idler photons are created and emitted collinearly by SPDC. The
remaining pump light is filtered by a silicon filter (Si). The
photons are coupled into singlemode optical fibres. They are then
separated and narrowly filtered (Filter, see text) before detected
by low-jitter single photon detectors (SPD).}} \label{setup}
\end{center}
\end{figure}

The spectral distribution of the downconverted light corresponds
to a coherence time of only $\tau_c=43\,fs$. In order to increase
this value, the photons have to be narrowly filtered. Such narrow
band filters are obtained by cascading two different types of
fibre Bragg gratings (see insert, figure~\ref{setup}). The first
one is a standard fibre Bragg grating (FBG$_s$) with a bandwidth
of $\sim 1\,nm$ and a high rejection rate ($>45\,dB$) over the
entire SPDC spectrum. This FBG reflects the desired wavelength,
requiring the use of a circulator. The second one is a
phase-shifted fibre Bragg grating (PSFBG$_s$), featuring a
10\,pm-wide transmission spectrum and a rejection window of only a
few nm. The downconverted light with a broad spectrum is sent into
port\,1 of the circulator. In port\,2, the FBG$_i$ reflects light
at $\lambda_s$ over $\sim1\,nm$ and transmits all the remaining
wavelengths. The reflected light then exits the circulator via
port\,3 and is further filtered by a PSFBG$_i$ ($\lambda_s$,
$\Delta\lambda_s = 10\,pm$). The remaining light, transmitted by
the FBG, is sent into another similar filter module centered at
$\lambda_i$.

The overall insertion loss of these filters (AOS GmbH) is 2-3\,dB.
The filters enable us to reduce the initial broad SPDC spectrum
down to a bandwidth of 10\,pm (1.2\,GHz), corresponding to 7\,cm
of coherence length in optical fibres, with a rejection of
$>$45\,dB over the whole SPDC spectrum. The filters are
temperature tuned and stabilized, allowing for a wavelength
precision of the order of 1\,pm over several weeks. Signal and
idler filters are placed at 1558\,nm and 1562\,nm respectively,
and independently tunable over 400\,pm, allowing for fine
adjustment.\footnote{In principle the idler-filter is not
necessary, but serves to improve the signal to noise ratio by
selecting the corresponding photon out of a broad spectrum of
non-correlated photons.}

\subsection{Source characteristics and applications}

This type of photon-pair source, combining a non-linear crystal
and narrow-band filters, can be pumped by either a CW or pulsed
laser. Here, the source is pumped by a CW diode laser with
external cavity (Toptica DL 100) at $\lambda_p=780\,nm$. The laser
is stabilized against the $D_2$ transition line of Rb\,87 allowing
for long term wavelength stability of less than 0.5\,pm.

An important parameter to characterize a light emitting source is
the spectral radiance $L_{\lambda}=hc^{2}\langle n
\rangle/\lambda^{5}$, with $h$ the Planck quantum, $c$ the speed
of light, $\lambda$ the photons wavelength and $\langle n \rangle$
the average number of created photons per mode
\cite{migdallRadiance}. Let us consider a single mode source, so
that $\langle n \rangle$ represents the average number of photons
per coherence time. If now, the bandwidth of the created spectrum
($\Delta \lambda_0$) is filtered to $\Delta \lambda_f$, on one
hand the number N of photons created per second is reduced by a
factor~ $r=\Delta\lambda_f/\Delta \lambda_0$ and on the other hand
the coherence time $\tau_c=0.44$ $\lambda^2/\Delta \lambda$ is
increased by r at the same time. This means that the number of
temporal modes per second M=$\tau_c^{-1}*s$ also diminishes by
$r$. Hence $\langle n \rangle= N/M$ (and consequently
$L_{\lambda}$) remains constant for any given source and is
independent on the spectral filtering. (for details see
\cite{klyshko,migdallRadiance}).

In order to avoid errors due to multiple photon emission, one is
limited to $\langle n \rangle \leq 1$ and in the case of a Poisson
photon distribution, this limit is on the order of 0.1
\cite{Scarani}. Our single mode source achieves a $\langle n
\rangle$ of 0.08~pairs per coherence length with only 7\,mW of
pump power injected into the wave\-guide. This radiance is one
order of magnitude larger than any comparable photon pair source
based on SPDC. Taking into account the losses due to absorption in
the crystal, fibre coupling and insertion losses in filters, an
overall optical transmission of $T=13\%$ per photon is obtained.
This leads to an effectively emitted spectral brightness
$E_{\lambda}$ into the single mode fibre of $3.9*10^5$ pairs
$pm^{-1} s^{-1}$.

This source can be adapted to match various applications
\cite{relay02, repeater, AHSPSfasel2004} which require either
maximal emission rates or long coherence length or any tradeoff in
between, simply by choosing a filter of suitable bandwidth. If,
for example, employed in conjunction with quantum memories, a
bandwidth on the order of up to 300\,MHz is required.

In principle, this is possible for any other source as well, but
in our case it is also practical since only a few mW of pump power
suffices to maintain the maximum $\langle n \rangle$. Furthermore,
due to the narrow bandwidth, chromatic dispersion as well as
polarization mode dispersion are negligible \cite{QKDfasel2004}.
In table~1, different approaches to entangled photon pair sources
are compared.

\Table{\label{tabl3} Comparison of different approaches to
entangled photon pair sources by the mean number of photons
$\langle n \rangle$ created per coherence time $\tau_c^{-1}$,
bandwidth $\Delta\lambda$ in $pm$, optical transmission $T$ in
including coupling efficiency into optical fibres in \%, and their
effectively emitted spectral brightness $E_{\lambda}$  per $s$ and
per $pm$. }\br \ns
Process&&$\langle n \rangle$&$\Delta\lambda$ & T&$E_{\lambda}$\\
&& [$\tau_c^{-1}]$&[$pm$]&[\%]&[$s^{-1} pm^{-1}]$\\
\mr
Filtered SPDC&&0.08& 10 &13&$3.9*10^5$ \\
4-wave-mixing$^a$& \cite{rarty07}& 0.025& 200&14&2*$10^4$ \\
Cavity SPDC& \cite{polzik} &0.012&0.02&14&$7.6*10^4$\\
Atomic ensemble$^b$& \cite{thompson} &0.02&0.01&35&$2.3*10^6$\\
Quantum dots$^{b\,c}$ &\cite{shields}&$N.A.$&620&8&$<1$\\
\br
\end{tabular}
\item[] $^{\rm a}$ Note that \cite{rarty07} works in pulsed mode,
such that $\langle n \rangle$ is given in pairs per pulse and
hence different from CW sources. \item[] $^{\rm b}$For quantum
dots and atomic ensembles, filtering induces additional losses due
to lack of energy correlation between the two photons of a pair
and hence their bandwidth is fixed. \item[] $^{\rm c}$ Coupling
efficiency for quantum dots is taken from an other experiment
\cite{yamamoto02}.
\end{indented}
\end{table}

From another point of view, our setup represents a heralded single
photon source at 1560\,nm with long coherence length and a
$P_1=0.13$. Up to now, most heralded single photon sources
\cite{thompson, heraldedKobayashi, heraldedSoujaeff,
heraldedTanzilli} have focused only on single photon and
multiphoton probabilities without taking into account the
bandwidth of the heralded photons. This figure is important since
chromatic dispersion in optical fibres will strongly reduce the
maximal communication distances \cite{AHSPSfasel2004}.

Given a Gaussian filter with a width of 10\,pm, the photons have a
corresponding coherence time of 350\,ps. This value is 7 times
greater than the resolution of state-of-the-art detectors, which
are described in the next section.

\section{Time resolution by detection}

In order to realize a complete asynchronous quantum communication
system we have implemented two free-running detectors based on
either nonlinear sum frequency generation (SFG) and Si-detectors,
or superconducting detectors. Si-APD detectors \cite{cova,
Rochas03a} offer both high quantum efficiencies (of ~50\%) and
temporal resolution better than 50\,ps, for wavelengths below
1$\,\mu$m. In order to detect light at 1560\,nm, photons can be
frequency converted to e.g. 600\,nm by SFG: A signal photon at
$1560\,nm$ is combined with a high power ($>200\,mW$) CW pump
laser at 980\,nm and then sent into a nonlinear PPLN WG crystal,
phase matched for up-conversion. Photons at 600\,nm are created
and detected with an overall quantum efficiency greater than
10\,\%, including coupling, losses and an APD detection efficiency
of 50\,\%. The dark count rate of such a setup can reach several
hundred kHz due to pump-dependent nonlinear noise \cite{robUp,
yamamotoUp}. However, one can reduce this noise level to a more
practical level, (which in turn reduces the upconversion quantum
efficiency) while maintaining the advantages of high temporal
resolution and passive detection. In this instance, we chose to
operate these detectors with 3\,\% detection efficiency and
30\,kHz noise.

\subsection{Superconducting detectors}

Alternatively, single photons can be detected by superconducting
sensors. One approach is given by transition edge detectors (TES)
\cite{TED}. These devices have a detection efficiency of over
80\,\%, very low counting rates (10\,kHz) and are not suitable for
time synchronization due to their poor time resolution (of the
order of 100\,ns).

Here we use another type of detector, referred as Superconducting
Single Photon Detector (SSPD), based on the so-called hotspot
process \cite{sspd2001}. A superconducting nanowire, made out of
ultrathin (3-5\,nm) Niobium Nitride (NbN) stripes of 100\,nm
width, is biased slightly below the critical current $I_c$. The
meander shaped nanowire (on a typical surface of 10$\mu$m*10$\mu
$m) is locally heated by the energy of an absorbed photon.
Subsequently, a normal hotspot is created. The nanowire section
available for the superconducting current is thus reduced and the
critical current density is locally exceeded. Hence the entire
stripe section becomes normal and an easily measurable voltage
pulse (mV) is generated before the NbN superconductivity is
restored \cite{sspd}.

The hotspot mechanism is a fast process with intrinsic response
times as low as 30\,ps \cite{sspd2000}, but practical devices are
limited by the kinetic inductance of the meander which is
typically hundreds of microns long \cite{sspd2006}. Hence, voltage
pulses are usually longer than 1\,ns and practical counting rates
are below 100\,MHz. The SSPD overall detection efficiency is
mainly limited by the poor absorption in the thin NbN layer in the
near infrared region, typically $20\%$ for 3-5\,nm thick films.

In our set-up the free running SSPD system (containing two
detectors) is cooled down to 1.7\,K by pumping a liquid Helium
bath to 3\,mbar. Both detectors are voltage biased and operated at
a superconducting current of 20\,$\mu$A corresponding to $90\%$ of
the critical current. The overall quantum efficiency at 1560\,nm
was measured to be $5\%$ and $5.5\%$ with dark count rates of
100\,Hz and 1\,kHz for the two different devices. The timing
jitter of the detection module including detector, electronics and
signal discrimination, has been established to be of the order of
70\,ps using a coincidence measurement with photon pairs of 40\,fs
coherence time.

\section{Experimental Results}

\subsection{Detection of high coherence photons.}

\begin{figure}[h!]
\begin{center}
\includegraphics[width=0.7\columnwidth]{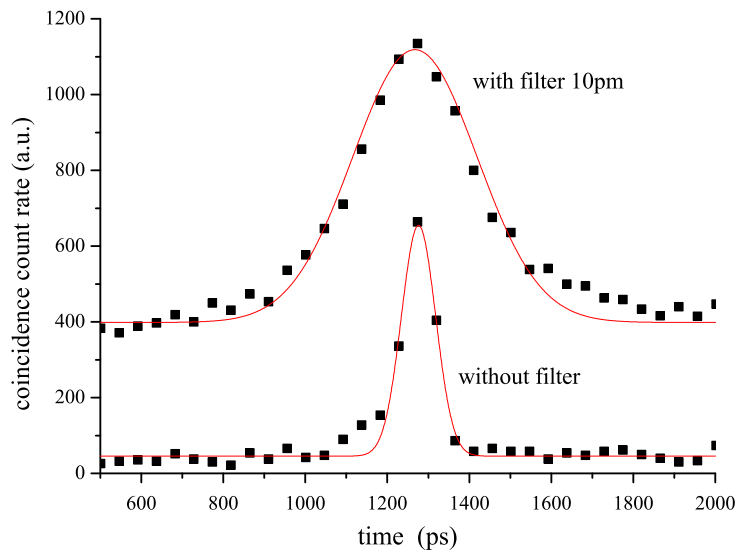}
\caption {\textit{Coincidence measurement of both, signal and
idler photons by a combination of up-conversion and Si-detectors.
If no filters are used (lower graph), the width of the curve
corresponds to the timing jitter of the detectors. For the case of
narrowly filtered photon pairs (upper graph), a broadening of the
response signal clearly shows the photon's coherence length
exceeding the detector's temporal resolution}}
\label{elargissement}
\end{center}
\end{figure}

The coherence length of filtered photon pairs can be measured with
a coincidence set-up (figure~\ref{setup}). Detecting the signal
photon ($\lambda_s$) at time $t_0$ projects the idler photon
($\lambda_s$) into the same temporal mode $t_0\pm \tau_c$ with an
uncertainty given by the photons coherence time $\tau_c$. In our
experiment, the detection signals are sent to a time to amplitude
converter (TAC) with a nominal temporal resolution of 45.5\,ps.
For the coincidence measurement of signal and idler photon, the
histogram of the time differences between the two electronic
signals is plotted in figure~\ref{elargissement}. Without
filtering, the SPDC photons coherence time is 40\,fs. The
coincidence peak (lower graph) with a measured FWHM of 80\,ps, is
the convolution of the two detectors signals and gives a precise
calibration of the setup resolution.

The second curve (upper graph) is obtained with the 10\,pm filters
inserted in the photon's path. In this case, $\tau_c$ is larger
than the detectors resolution and the coincidence measurement
shows a temporal distribution governed by the photons coherence
time. We observe a significant widening of the coincidence peak
(400\,ps), corresponding to a deconvolved width of 285\,ps for
each photon. The measured value is less than the theoretical
350\,ps due to the filters non-gaussian spectrum and slightly
larger bandwidth, than specified. This measurement proves that we
have, on one hand, created a photon pair source based on SPDC with
highly coherent photons, and on the other hand, demonstrated that
recent photon detectors with high temporal resolution are capable
of resolving the photons arrival time with sub-coherence time
precision.

\subsection{A Hong-Ou-Mandel experiment}

In order to demonstrate the performance of our system we conduct a
HOM experiment. In such an experiment, two indistinguishable
photons superposed on a 50/50 beamsplitter (BS) bunch together
into the same output mode due to their bosonic nature
\cite{HOM87}. Indistinguishability requires, besides identical
polarization, spectral and spatial modes, that the photons enter
the BS simultaneously. This is normally obtained by using photons
originating from the same pair or synchronized pulsed emission
from different sources.

\begin{figure}[h!]
\begin{center}
\includegraphics[width=0.7\columnwidth]{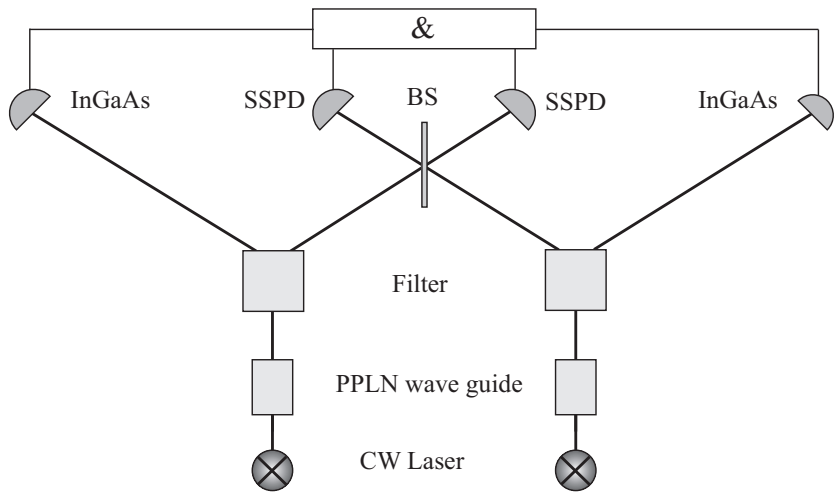}
\caption{\textit{ Experimental setup for a CW HOM-dip experiment.
Pairs of photons are independently created by autonomous
unsynchronized CW sources. The photons are filtered to 10\,pm
(Filter) in order to increase their coherence time to ~300\,ps.
One photon from each source is sent onto a 50/50 beamsplitter (BS)
and coincidence count rates between the two high resolution
superconducting detectors (SSPD) at the output ports are recorded.
Further coincidence detections ($\&$) by the two InGaAs APD ensure
that we only observe events with photon pairs from different
sources.}} \label{setupDip}
\end{center}
\end{figure}

In the case of independent sources with CW excitation, due to the
lack of any synchronization, one has to postselect photon pairs
arriving at the same time at the beamsplitter by detection
\cite{halderSwap}. This requires detectors with a temporal
resolution superior to the coherence time of the photons. The two
output modes are each connected to a SSPD as shown in
figure~\ref{setupDip}, and for the case where both SSPDs click,
the arrival time difference $\tau$ of the photons is recorded by a
time to digital converter (TDC), connected to a computer.

\begin{figure}[h!]
\begin{center}
\includegraphics[width=0.7\columnwidth]{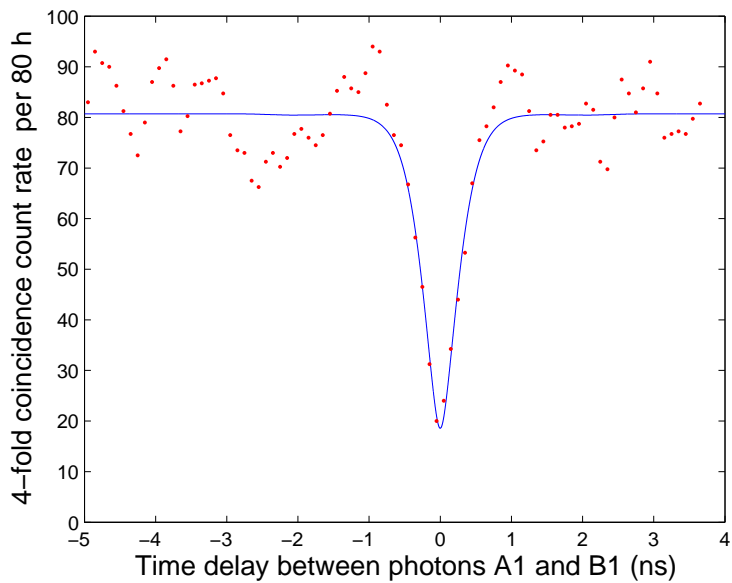}
\caption{\textit{ Coincidence count rate between the two SSPD as a
function of the measured temporal delay $\tau$ of two identical
photons impinging on a beamsplitter and originating from
independent sources. For $\tau=0$ a HOM-dip with a visibility of
78\% is observed. }} \label{dip}
\end{center}
\end{figure}

If this is repeated continuously and the coincidence count rate
$R$ of the two detectors is plotted as a function of the photons
relative time delay $\tau$ (figure~\ref{dip}), a decrease in $R$
is observed for $\tau=0$, due to photon bunching, giving rise to a
``dip". In order to ensure that the two signal photons originate
from different sources, the related idler photons of each pair
must also be detected, as illustrated in figure~\ref{setupDip}. A
raw dip visibility -defined as $(V_{max}-V_{min}/(V_{max})$- of
78\% is observed, which proves the ability to temporally resolve
the photons' arrival times at the BS. We measure approximately one
4-fold coincidence per time slot of the TDC (45.5\,ps) for each
hour of measurement.

Note that we detect all possible arrival-time differences in
parallel and each of the time differences is given by the
arbitrary emission times of the CW sources. (This corresponds to a
4-fold coincidence count rate of 400 events per hour for a
temporal range of $\pm$ 10\,ns). This value is limited by the
probability of two photons, being emitted independently and
arbitrarily, to arrive at a given time at BS. The final 4-fold
coincidence count rate is further affected by losses in fibre
coupling, absorption in the filters and low detector efficiencies.

The detectors' temporal resolution is sufficient to resolve the
photons coherence length. The limited visibility is predominantly
due to multiphoton creation. Thus the visibility could be further
increased by reducing the pump power, with the consequence of
longer measurement times.

\section{Conclusion}

In conclusion we have realized a set-up of two autonomous CW
photon pair sources with long coherence times in combination with
new-generation high resolution detectors. A spectral radiance
$L_{\lambda}$ corresponding to 0.08 photons per coherence length
and an emitted spectral brightness $E_\lambda$ of $3.9*10^5$ pairs
$s^{-1} pm^{-1}$ are achieved. Such sources can be used in various
quantum communication protocols without any need for pulse

synchronization as in field experiments for long distance quantum
communication. Furthermore their bandwidth is compatible with
quantum memories. Even though some efforts have still to be made
by both sides, source and memory bandwidths are within the same
order of magnitude and hence make asynchronous quantum repeaters
\cite{mmm} more feasible.

\ack We acknowledge technical support by J.-D. Gautier and C.
Barreiro. This work was supported by the EU projects QAP and
SINPHONIA and by the Swiss NCCR Quantum Photonics.

\section*{References}
\bibliography{H:/BIB/thesis}
\bibliographystyle{unsrt}

\end{document}